# Determining the nature of quantum resonances by probing elastic and reactive scattering in cold collisions

Prerna Paliwal[1†], Nabanita Deb[1†], Daniel M. Reich[2], Ad van der Avoird[3], Christiane P. Koch[2]*, Edvardas Narevicius[1]*

[1]Department of Chemical and Biological Physics, Weizmann Institute of Science, Rehovot 76100, Israel
[2]Dahlem Center for Complex Quantum Systems and Fachbereich Physik, Freie Universität Berlin, Arnimallee 14, 14195 Berlin, Germany
[3]Institute of Theoretical Chemistry, Institute for Molecules and Materials, Radboud University, Heyendaalseweg 135, 6525 AJ, Nijmegen, Netherlands
*Correspondence to E. Narevicius <edvardas.narevicius@weizmann.ac.il>, C. P. Koch <christiane.koch@fu-berlin.de>
[†]These authors contributed equally

**Abstract: Scattering resonances play a central role in collision processes in physics and chemistry. They help building an intuitive understanding of the collision dynamics due to the spatial localization of the scattering wavefunctions. For resonances that are localized in the reaction region, located at short separation behind the centrifugal barrier, sharp peaks in the reaction rates are the characteristic signature, observed recently with state-of-the-art experiments in low energy collisions. If, however, the localization occurs outside of the reaction region, mostly the elastic scattering is modified. This may occur due to above barrier resonances, the quantum analogue of classical orbiting. By probing both elastic and inelastic scattering of metastable helium with deuterium molecules in merged beam experiments, we differentiate between the nature of quantum resonances – tunneling vs above barrier – and corroborate our findings by calculating the corresponding scattering wavefunctions.**

A scattering resonance, or temporary low energy "collision complex", can be formed by various mechanisms[1,2]. Trapping of the colliding particles can occur due to the transfer of relative kinetic energy to other degrees of freedom, giving rise to Feshbach resonances[3,4]. These are nowadays routinely used to tune interactions, for example between cold atoms[5] or atoms and molecules[6]. Feshbach resonances arise from the coupling of scattering states to a bound state belonging to another scattering channel which is asymptotically characterized by a different set of quantum numbers. The coupling to a bound state results in a resonance wavefunction that is localized at a short interaction range. On the other hand, shape and orbiting resonances form on a single potential curve which possesses a barrier. While shape resonances are typically associated with tunneling through the barrier, a resonance may also arise due to quantum reflection above the barrier. We refer to the resonances formed on or above the barrier as orbiting resonances, due to their analogy with classical orbiting. This distinction is important since the wavefunction of an orbiting resonance, in contrast to those of Feshbach and shape resonances, is highly de-localized and the collision complex thus short-lived. In reactions and inelastic processes, which can be described by capture models separating long- and short-range dynamics[7,8], a resonance state localized at short separation behind the centrifugal barrier will have the strongest effect. In contrast, a resonance that "hovers" above the centrifugal barrier will be more relevant for elastic scattering. As such, investigation of both elastic and short-range dominated inelastic channels is a convenient way to probe the localization of the resonance wavefunction and unequivocally identify the different underlying physical mechanisms.

In order to observe quantum resonances in the scattering experiments, one needs to reach collision energies ($E/k_B$) corresponding to a few kelvin. Such scattering resonances were first identified in pioneering experiments performed by Scoles and co-workers[9] with resonances clearly resolved by Toennies and co-workers[10–13], where a crossed beam set-up was used to achieve collision energies corresponding to 60 K and 5 K respectively. Boesten et al. observed a shape resonance in cold atom scattering by pulsed photoassociation[14]. Dynamical Feshbach resonances were reported in the benchmark reaction, F + H₂ isotopologues, in crossed beam experiments by Skodje et al.[15] and Qiu et al.[16], at collision energies going down to 100 K. In a different approach, quantum resonances have also been identified in anion photoelectron spectroscopy by directly probing the transition state dynamics[17,18]. With the recent success in reaching sub-kelvin collision energies using merged beams[19,20], shape resonances have also been detected in Penning Ionization at a few millikelvin. Chefdeville et al. used a crossed beam set-up to observe scattering resonances in inelastic molecular scattering at collision energies down to 5 K[21]. Vogels et al. imaged inelastic scattering resonances using the velocity map imaging technique together with a DC-electric field decelerator[22,23]. Bergeat et al. studied inelastic scattering observing resonances in the spin-orbit excitation of atomic carbon[24]. Low-lying shape resonances have also been observed using pulsed-field-ionization zero-kinetic-energy photoelectron spectroscopy in half-collisions of $H^+ - H$ and $D^+ - D$.[25]

Resonances observed in all previous experiments have been identified by observing peaks in the integrated cross section, partial integrated cross section[23] or backward scattering spectrum[26] measured as a function of collision energy. So far, the nature of the resonances could only be inferred using theoretical analysis. Orbiting and shape type resonances may occur in both elastic and inelastic collisions, however a different formation mechanism leads to sensitivity towards the range of interaction. In order to differentiate shape from orbiting resonances experimentally one needs to be able to probe the localization of the resonance wavefunctions. Here, we investigate two different processes for the same collision system that are sensitive to either the long-range or the short-range of the interaction potential to allow for this differentiation. In our experiments, we identify scattering resonances in the elastic scattering (I) that probes the entire interaction range and compare it with previously studied Penning ionization[27] (II) rate of metastable helium (He*) with D₂ which is mainly sensitive to the short-range interaction.

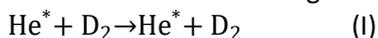
$He^* + D_2 \rightarrow He^* + D_2$ (I)
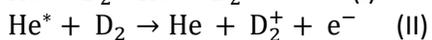
$He^* + D_2 \rightarrow He + D_2^+ + e^-$ (II)

Detecting and imaging low energy resonances in the elastic scattering channel at a few kelvin is particularly challenging due to the small scattering angle of the products in the laboratory frame of reference. Here, we use zero-energy electron recoil assisted velocity map imaging together with merged beam technique to investigate low energy elastic collisions. Here, we are able to image above barrier quantum resonances and elucidate their effect on observable quantities distinguishing them from tunneling resonances. We show that the reactive scattering is far more sensitive to tunneling resonances that are localized at short internuclear distances, whereas both tunneling and quantum reflection lead to formation of additional resonance structures detected in the elastic scattering process.

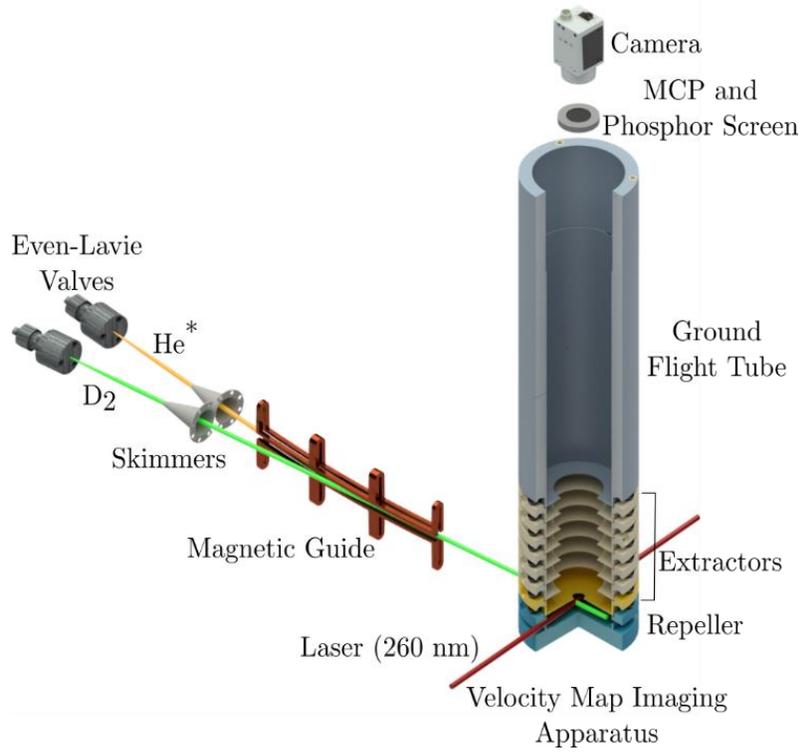

**Fig. 1 | Experimental set-up.** Supersonic beams of $D_2$ and metastable helium ($He^*$) are produced by two pulsed Even–Lavie valves and merged using a magnetic guide. A 260 nm laser is then used to single photon ionize $He^*$ atoms that are subsequently detected via velocity map imaging set-up

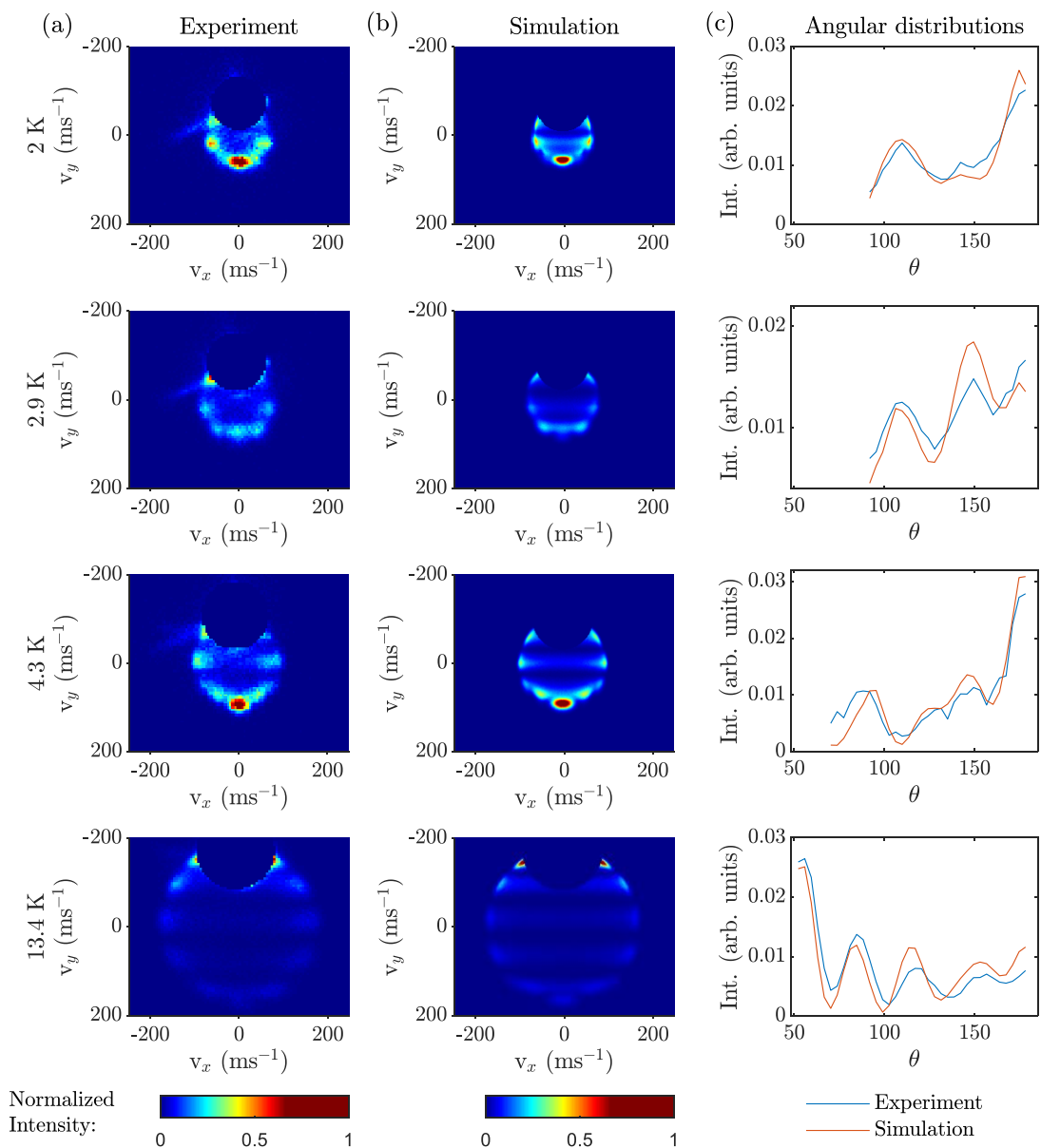

**Fig. 2 | VMI images and angular distributions.** Each column (left to right) refers to experimental VMI images, simulated VMI images and the angular distributions as a function of scattering angle, $\theta$, obtained from experiments (blue) and simulations (red), respectively. Each row (top to bottom) corresponds to data obtained for collision energies ($E/k_B$) of 2.0±0.1 K, 2.9±0.2 K, 4.3±0.2 K and 13.4±0.5 K respectively, the error bars are for experimental images only and indicate standard deviation. The *X* and *Y*-axis refers to the velocity of He* in the center-of-mass frame. The *Y*-axis is the direction of the relative velocity vector, with the forward direction ($\theta = 0°$) pointing up in all the images. To bring all the images to the same color scale, the images are normalized (discussion in Methods).

**Results and Discussion:**

The key to our ability to resolve resonances in the elastic differential cross section (DCS) is the combination of the merged beam technique[19] with velocity map imaging (VMI)[28] detection assisted by near-threshold photo-ionization and a low velocity spread of the reactants. The apparatus used in this study is illustrated in Fig. 1 and described in Methods. We investigate the low energy elastic scattering between helium atoms in the $2\,^3S_1$ metastable state (He*) and normal-$D_2$ (2/3 ortho-$D_2$ with $j = 0$ and 1/3 para-$D_2$ with $j = 1$, where $j$ is the rotational quantum number). In our experiment, we probe elastically scattered He* atoms by single-photon near-threshold ionization at 260 nm. This ensures that the electrons carry near-zero kinetic energy thus avoiding image blurring due to the photo-electron recoil effect[29]. Another necessary condition for the measurement of elastic scattering DCS is a narrow velocity spread of the He* beam. Our supersonic source, the Even-Lavie valve[30] combined with a dielectric barrier discharge (DBD) generates a 150 mK cold beam of He* localizing the unscattered part of atoms to a small area on our detector. Note that in the collision energy range of our experiment, no internal state-changing collisions are possible; excitations are energetically not allowed and quenching is also not possible because the $D_2$ molecules produced by the supersonic expansion are already in their lowest ro-vibrational state.

We present in Fig. 2 the VMI images of the scattered He* obtained experimentally and simulated theoretically at different relative velocities corresponding to collision energies ($E/k_B$) of 2.0±0.1 K, 2.9±0.2 K, 4.3±0.2 K and 13.4±0.5 K, with the error bars indicating the standard deviation. We have removed the forward part of the measured image which is dominated by the direct He* beam that does not undergo scattering. For clarity and comparison, we have removed the corresponding area also from the simulated images. Since, no inelastic scattering is possible at our experimental collision energies and the total energy is conserved for elastic collisions, the images show a single scattering ring corresponding to a radius proportional to the relative velocity. We also observe diffraction oscillations in all the VMI images due to interference between different partial waves given by integer quantum number $l$. The maximum number of interfering partial waves is determined by collision energy. The bands obtained in the VMI images are a result of projecting the 3D sphere in momentum space onto a 2D microchannel plate (MCP) detector. The angular distributions reflect the differential cross sections of the collision process. The agreement between the calculated and measured angular distributions visible in Fig. 2 is gratifying and attests once more to the accuracy of the potential energy surface[31].

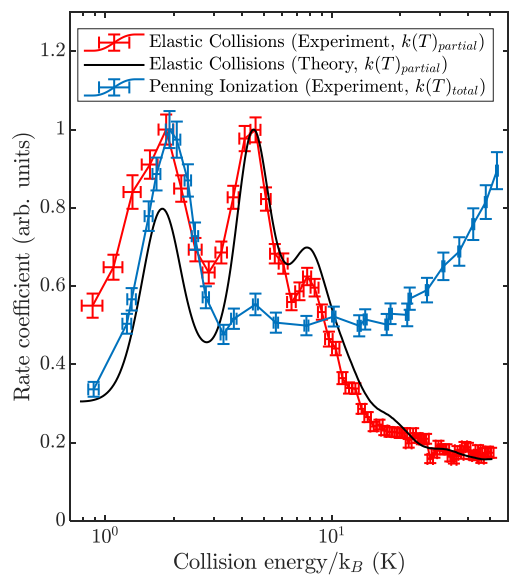

**Fig. 3 | Relative rate coefficients ($k(T)$) vs collision energy/k$_B$ (K), for elastic scattering and Penning ionization of He$^*$ − D$_2$.** The red data points with errorbars denote the experimentally measured partial relative rate coefficient for the elastic scattering of He$^*$ − D$_2$ where the scattering signal is integrated over the backward hemisphere. The black curve is the theoretically predicted partial relative rate coefficient for elastic scattering including full kinematics of the experiment (see Methods). The blue data points with errorbars represent the experimentally measured relative rate coefficient for Penning ionization of the same system (from Ref. 27). The error bars represent standard deviation in collision energy and rate coefficients. The red and blue lines are just joining the experimental data points. The relative scaling of the red and blue curve is arbitrary and is made equal to one at the energy corresponding to the first maxima of both curves.

In addition to the diffraction oscillations, the backward scattering is seen to be strongly enhanced in Fig. 2 at collision energies ($E/k_B$) of 2.0 K and 4.3 K. As the collision energy approaches the resonant energy, a single partial wave ($l_{res}$) structure, dictated by square of the Legendre polynomial, $\left|P_{l_{res}}(cos\theta)\right|^2$, starts to dominate the angular distribution. This leads to an enhancement of forward-backward scattering and has also been reported for other systems[22,23,32–34]. In order to find the energy positions of these states, we measure the partial relative rate coefficient for the elastic collision between He$^*$−D$_2$ as a function of collision energy. It is illustrated by the red curve in Fig. 3 for the collision energy range 0.9 − 50 K. For a given energy, the partial rate coefficient is obtained by counting the number of He$^*$ atoms scattered in the backward hemisphere (θ ≥ 90°) of the image and scaling it by the product of number of He$^*$ in the reactant beam and D$_2$ beam intensity (see Methods). The black curve in Fig. 3 is the theoretically predicted partial rate coefficient, which is calculated by taking into account full kinematics of the experiment. We obtain theoretical partial cross section by performing close-coupling quantum scattering calculations for collisions with ortho-D$_2$ ($j = 0$) and para-D$_2$ ($j = 1$) using the theoretical framework established in Klein *et al.*[31] and discussed in Methods. A comparison of theoretically obtained partial and total rate coefficients for elastic scattering of He$^*$ with normal- D$_2$ is shown in Supplementary Fig. 1. The experimentally obtained partial rate coefficient in Fig. 3 shows good agreement with the theoretically predicted partial rate coefficient for the same energy range.

Additionally, the blue curve in Fig. 3 shows the rate coefficient obtained in previously performed Penning ionization experiments[27]. The experimental curve shows a possible feature, predicted by theory[27] at 4.6 K, with its amplitude contained within the statistical error. However, in case of elastic collisions at the same energy this feature becomes a fully resolved resonance peak with the amplitude comparable to the low-energy resonance at 1.8 K. In addition, a new minor resonance peak appears at 7.8 K for elastic collisions. This indicates a different physical mechanism that is responsible for the formation of these two resonance states as compared to the major shape resonance observed in the reactive Penning ionization process.

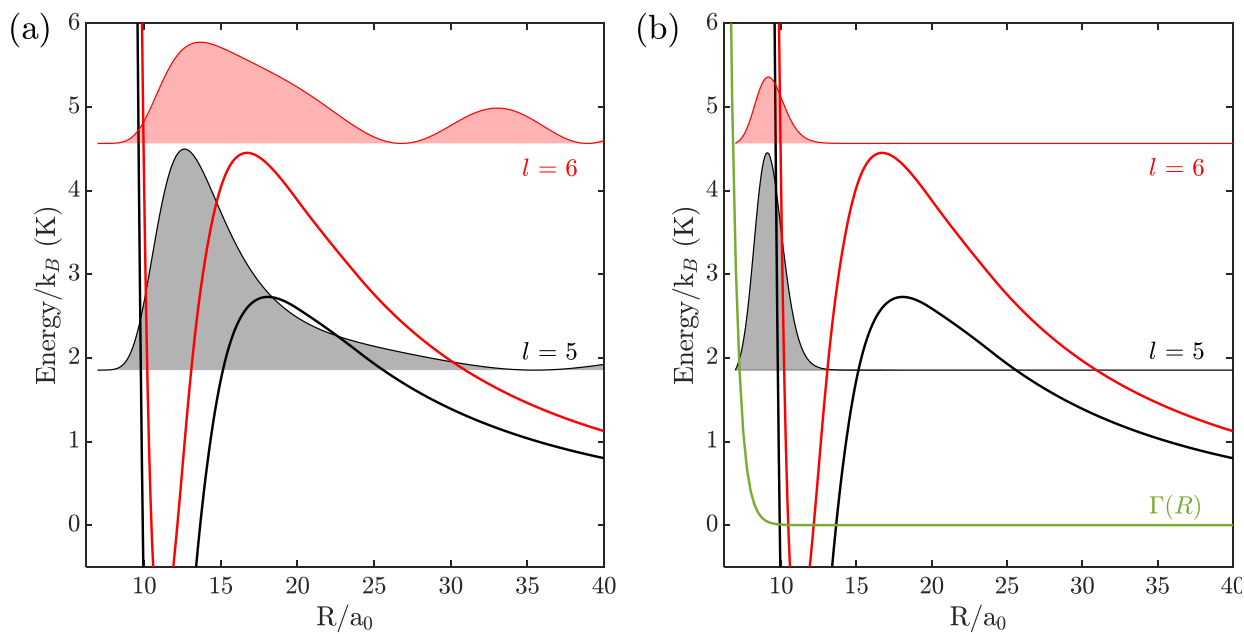

**Fig. 4 | Partial wave analysis.** The effective isotropic potential curves are presented as a function of the distance between $He^* - D_2$ together with the (a) corresponding resonance wavefunctions squared and (b) the resonance wavefunctions squared multiplied by the Penning ionization rate $\Gamma(R)/\hbar$ shown in green. Black and red denote the partial wave channels $l$ = 5 and 6, respectively. The shading illustrates the probability density for elastic scattering (a) and Penning ionization (b).

In order to confirm the character of these resonances, we perform partial wave analysis, where we calculate the resonance energies using complex absorbing potential method[35] and the resonance wavefunctions using Numerov algorithm. For simplicity, we consider only the isotropic part of the potential in this analysis. The probability density of the colliding particles, given by $|\psi|^2$, as a function of interparticle separation is indicated by the shaded regions in Fig. 4(a) for the two resonances at 1.8 K and 4.4 K, corresponding to the $l$ = 5 (black) and $l$ = 6 (red) partial wave channels, respectively. The energy of the $l$ = 5 resonance lies below the centrifugal barrier such that the amplitude is trapped behind the barrier — this is a shape resonance. In contrast, the energy of the $l$ = 6 resonance lies well above the peak of the centrifugal barrier (the same is true for the $l$ = 7 resonance). These are above the barrier orbiting resonances. A more rigorous analysis for identifying the contributions of different $j$ and $l$ to individual resonances is done by calculating rotational constants as illustrated in Fig. 5 and discussed in Methods. The resonances at 1.8±0.1 K, 4.4±0.2 K and 7.8±0.3 K are indeed formed with the contribution of a single partial wave, $l$ = 5, 6 and 7 respectively, which dominate over the background. Due to different multiplicities in $j$, several $J$ values contribute to a single resonance in the case of He* colliding with para $D_2$, where $J$ is the total angular momentum. This is in contrast to He* colliding with ortho $D_2$, where $j = 0$ and only a single $J$ contributes (note that the value of $J$ ranges from $|j − l|$ to $|j + l|$). However, in the experiment, these structures are only visible as a single convoluted peak.

To explain, why the $l$ = 6 and $l$ = 7 resonances were not observed in the Penning ionization reactions, we calculate the Penning ionization probability as a function of internuclear distance. It is obtained by scaling the probability density $|\psi|^2$ by the ionization rate $\Gamma(R)/\hbar$, approximated by a single exponential term and shown by the green curve in Fig. 4(b), where the $\Gamma(R)$ is taken from Yun et al[36]. Clearly, the Penning ionization probability ($|\psi|^2\Gamma(R)$) of the $l$ = 6 resonance is much smaller than that for $l$ = 5, due to the very strong weight that the ionization rate gives to the probability amplitude at the shortest distances. While the $l$ = 6 above barrier resonance show some localization in its wavefunction as shown in Fig. 4(b) which is in agreement with the theoretical prediction for Penning ionization[27], it is not strong enough to considerably enhance Penning ionization, as is obvious from the blue curve in Fig. 3 (the same is true for $l$ = 7).

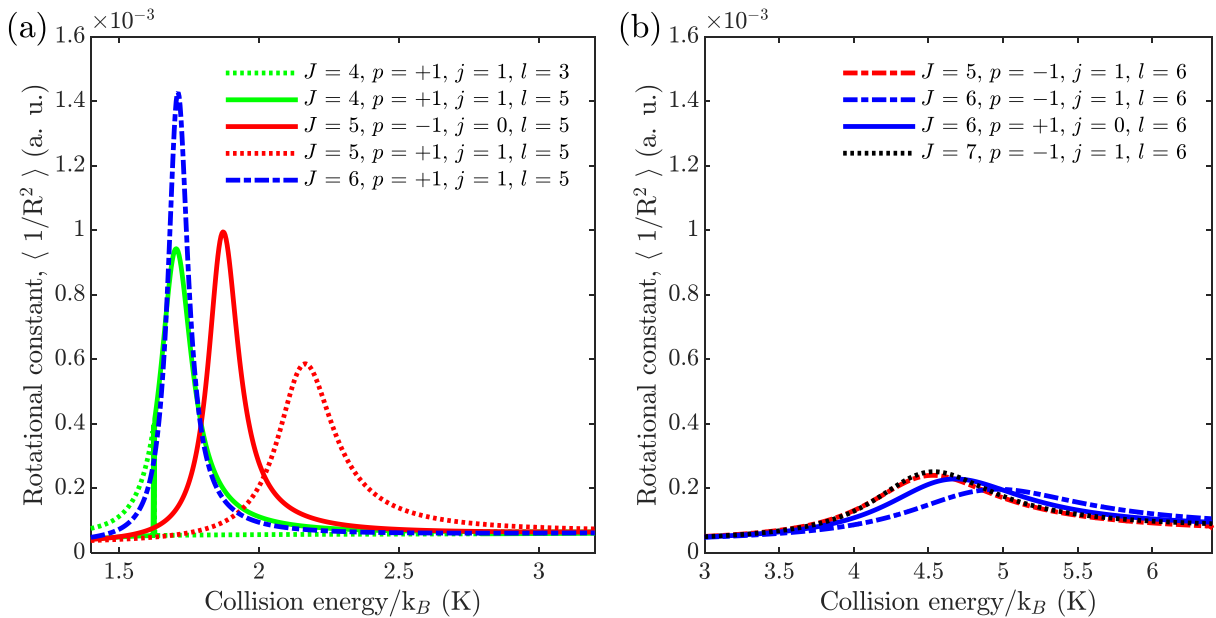

**Fig. 5 | Characterization of resonances.** Expectation value of the rotational constant $\langle 1/R^2 \rangle$ as a function of eigen state energy for ortho-$D_2$ ($j = 0$) and para-$D_2$ ($j = 1$) at (a) 1.8±0.1 K and (b) 4.4±0.2 K. Here, $J, p, j$ and $l$ denote total angular momentum, parity, rotational quantum number and partial wave respectively.

The combination of two state-of-the-art techniques, merged beams and high-resolution velocity-map imaging, enables us to observe the resonances in the elastic collisions of He* with $D_2$ at collision energies going down to a kelvin, where two new resonance structures are revealed as compared to previously studied Penning ionization reaction. This suggests a different resonance formation mechanism in play. The dynamics of Penning ionization process can be naturally divided into the long-rage part that is governed by centrifugal barriers and the short-range part that defines the reaction probability. This is a hallmark of a process that can be conveniently described by capture models[37]. As such, we show that by comparing the elastic scattering process, that is sensitive to the whole internuclear interaction range, with highly localized Penning ionization process we can unambiguously differentiate between localized shape type resonances and short-lived orbiting ones. We expect this approach to be general and applicable to other processes described by capture models.

**Acknowledgements:** We acknowledge financial support from the European Research Council and the Israel Science Foundation. Additional financial support from the German-Israeli Foundation, grant no. 1254, is gratefully acknowledged. C.P.K. is grateful for a Rosi and Max Varon Visiting Professorship. Correspondence and requests for materials should be addressed to E.N. and C.P.K.

**Methods**:

**Experiment:** Supersonic beams of He and $D_2$ are produced by adiabatically expanding neat gases at backing pressures of 35 bars via two pulsed (10 Hz) and cooled Even–Lavie valves[30]. The $D_2$ beam is produced with a mean velocity ranging from 990−1550 m/s (speed-ratio ~30), generated by changing the temperature of the valve, and this is how the collision energy is tuned from 0.9 K to 50 K (by changing the relative velocity between the two beams, keeping the He* beam velocity constant). The measurement is carried out with normal $D_2$, consisting of 2/3 part of ortho and 1/3 of para. The supersonic expansion used in the production of the beams creates $D_2$ in the lowest $j$ states of its ortho and para components, resulting in two third of population in $j = 0$ and a third in $j = 1$. The He beam, produced with a mean velocity 906 m/s (speed-ratio 70), is excited to the paramagnetic $2^3S_1$ metastable state (He*) by a DBD[38] located at the valve orifice. Only the low-field-seeking Zeeman sublevel, $m_j = 1$, is confined in the two dimensions during the transit through the 20-cm long magnetic guide. The valves are timed such that the He* and the $D_2$ beams arrive at the laser ionization volume (detection zone) located at the center of the VMI set-up at the same time. Both the beams pass through 4-mm-diameter skimmers located 10 cm after the valve orifice. The $D_2$ beam then travels straight to the reaction chamber, whereas the He* beam enters the magnetic guide, which has a 10° curve, and is subsequently merged with the straight-propagating $D_2$ beam. The beams pass through a 3 mm aperture located at the entrance of the VMI set-up attached perpendicular to the beam propagation axis. The electrostatic lens in the VMI set-up consists of eight separate plates maintained at constant voltages throughout the measurement, followed by a 9th grounded plate. The first extractor plate placed immediately above the repeller plate has an aperture of 1 cm in diameter. The remaining seven plates have an aperture of 4 cm diameter. The voltages are distributed linearly from 2000 V on the repeller plate to 1160 V on the eighth plate in order to focus products on the MCP detector plane located after the ~1 m long grounded flight tube as shown in Fig. 1. A 260 nm laser (~10 µJ), obtained by doubling the output of an OPO laser, pumped by 355 nm pulsed (10 Hz) Nd:YAG source, is used to single photon ionize the He*. The laser propagation axis is perpendicular to both the beam propagation axis and the VMI set-up. He$^+$ ions thus formed are accelerated towards the phosphor screen located behind the VMI plates and MCP. The fluorescence generated from the electrons (formed from He$^+$ hitting the MCP) as it impinges on the phosphor screen is then imaged on the camera. A real time 'centroiding' is performed to determine the *X-Y* coordinate of each event on the MCP, thus correlating an electrical signal to the position and number of ions. The MCP is time-gated for detecting mass/charge of 4 amu/e by applying a 50 ns high voltage pulse in order to reduce the undesired ions from associative ionization and any other processes such as ionization of background gases by laser and He*. However, the collisions of He* with $D_2$ also produce Penning ionization product, $D_2^+$ which has the same mass/charge as that of He$^+$. Note that Penning ionization products can also be detected via VMI[39,40], nevertheless, their absence in our measurements is explained by several factors. The rate of Penning ionization between He* and $D_2$ is at least 70 times slower (obtained from theoretical cross sections) when compared to partial rate coefficient (integrated over backward hemisphere) in case of elastic collisions. $D_2^+$ ions formed in the PI reaction are continuously extracted over period of 10 μs. $He^+$ ions that are detected after photoionization arrive within 50 ns time window achieved by MCP gating. This provides another signal to background $D_2^+$ ion ratio of 200 bringing the total signal to background ratio to 70×200=14000. Thus, the detection of $D_2^+$ is reduced by more than four orders of magnitude as compared to $He^+$ and does not interfere with our measurements.

The VMI apparatus is calibrated by accumulating images at different relative velocities and then determining the shift in the center of images obtained, here the center of the images corresponds to the center-of-mass velocity of He*- $D_2$ pair. The velocity per camera pixel is determined to be 10.9 m/s for $He^+$. The experimental VMI images shown in Fig. 2 are obtained by binning the *X-Y* coordinates of each MCP event onto a 2D array with a mesh spacing equal to 0.5 camera pixel. These images are normalized by dividing each element of the 2D array by the total intensity in respective backward hemispheres and subsequently multiplying by the partial rate coefficient for that energy. All images are represented on the same colorscale where the maximum is set to the highest intensity for image corresponding to 1.8 K. The angular distribution for the images is evaluated by integrating the intensity in an annulus whose edges are determined by the width of the intense ring visible in all VMI images. Here, the angular grid spacing is taken to be 3.6° inside the annulus.

We characterize the He* beam by counting the number of He* atoms present in the same ionization region in the absence of the $D_2$ beam. This is achieved by defocusing the $He^+$ signal on the VMI camera after single photon ionization using 260 nm laser. Since more than 10 ions are present in each ionization event, we defocus the $He^+$ ions by changing the voltage on the first extractor plate of our VMI apparatus, increasing the spot size to a dimension that allows convenient separation between individual ions. A separate MCP placed on-axis with the direction of propagation of the beams is also used to characterize the He* beam. The $D_2$ beam is characterized by a time-of-flight mass spectrometer (TOF-MS)[41] positioned perpendicular to the beam propagation direction, using an ionization filament to form $D_2^+$. The TOF-MS is used in multipulsed mode with 10 us intervals to obtain the longitudinal profile of the beam. At any given collision energy, the partial rate coefficient is obtained by scaling the number of He* atoms scattered in the backward hemisphere of the VMI image and collected for 21000 laser shots by the product of number of He* in the reactant beam and $D_2$ beam intensity, averaged over 1500 laser shots and 100 shots respectively.

**Theory:** To obtain cross sections and rate coefficients, close-coupling quantum scattering calculations have been performed for collisions with ortho-$D_2$ ($j = 0$) and para-$D_2$ ($j = 1$), using the theoretical framework established in Klein *et al.*[31] The angular basis functions are constructed as products of the rotational state of the $D_2$ molecule and the partial wave describing the orbital motion of the whole complex and symmetry-adapted to the values of the total angular momentum *J* and spectroscopic parity. The basis set is restricted to even (odd) rotational states of the dimer for collisions with ortho-$D_2$ (para-$D_2$). Differential elastic and inelastic scattering cross sections for energies up to 50 K are obtained using a basis set including all of the relevant functions with $J \leq 30$. This converges our results to an accuracy of below 1% even for the highest scattering energies considered. The scattering wavefunctions are obtained by propagating from $7.0a_0$ to $200.0a_0$ by means of a renormalized Numerov propagator. Short-range and asymptotic boundary conditions are imposed afterwards, as described in Janssen *et al*[42]. The final cross section used in the simulations (described below in Simulation details) are the weighted average of the ortho and para-$D_2$ cross sections.

Scattering resonances are identified by diagonalization of the nuclear Hamiltonian describing collisions between metastable helium in the $2\,^3S_1$ state and molecular deuterium for fixed total angular momentum and spectroscopic parity. The corresponding potential energy surface is adopted from Klein *et al.*[31], where it had been used to obtain resonance positions and the overall behavior of rate coefficient for the electronically identical He$^*$ — H$_2$ system, yielding excellent agreement between theoretical results and measured data. Here, for the He$^*$ — D$_2$ case, a mapped grid using 6144 grid points and ranging from $2.0 a_0$ to $20000.0 a_0$ is employed for the diagonalization, in conjunction with a Fourier-basis representation of the kinetic energy operator and a transmission-free complex absorbing potential at the outer boundary[43]. Resonances are identified as local maxima of the expectation value of the rotational constant $\langle 1/R^2 \rangle$ as a function of energy[44], which is illustrated in Fig. 5 for the resonances at 1.8±0.1 K and 4.4±0.2 K. These clearly have a predominantly $l$ = 5 and $l$ = 6 character respectively. Since only $J$ is a good quantum number, the scattering states may have contributions from several $l$. We take a scattering state to be of $l$-character if 50% or more of its population resides in that $l$-channel. For all $J$ quantum numbers contributing to the resonance around 4.4 K, we observe $l$ = 6 character for their full energetic width. The same behavior can be observed for the resonance with $l$ = 5 character around 1.8 K with a minor exception: for $J$ = 4 and positive parity the character of the resonance flips from $l$ = 3 (dashed green curve in Fig. 5) to $l$ = 5 (solid green curve in Fig. 5) at the resonance's left tail at about 1.62 K. This slight anomaly can also be observed when looking at the diabatic population in the $l$ = 5 state, which is only slightly larger than 60% at the resonance peak. Conversely, the diabatic population in the $l$ = 5, respectively $l$ = 6, states for all other cases we show in Fig. 5 is at or above 95%. This indicates that the anisotropy-induced coupling between the diabatic $l$ surfaces is larger for the $J$ = 4 He$^*$ — para D$_2$ resonance than in all other cases we have investigated.

**Simulation details:** The detection method employed in our experiment measures the partial relative rate coefficient and the angular distribution of the elastically scattered products. In order to convert the theoretical cross sections into experimental observables, we have developed a computer program to simulate the He$^*$ — D$_2$ elastic scattering, taking into consideration full kinematics of the experiment.

In the simulation, the two supersonic beams are assumed to have a Gaussian profile in velocity and temporal space. The mean velocities and the velocity spreads of the beams are known from the beam characterization described in the experimental section of Methods. However, the temporal spread of the beams are not known, so we have optimized this parameter to achieve best fit to the experiment. In the experiment, the time difference between the valves is adjusted such that the beams exactly overlap only at the center of the VMI set-up, however, the interaction region starts at the exit of the magnetic guide (where the tail ends of the Gaussian beams start overlapping) and continues all the way upto the detection zone. Accordingly, in the simulation, the initial interaction time ($t_0$) is defined as the time when the He$^*$ beam exits the magnetic guide and the final interaction time ($t_{laser}$), as the time when the laser is fired. All collisions taking place between time $t_0$ and $t_{laser}$ are considered. For any given time $t$ ($t_0 \leq t \leq t_{laser}$), the number of scattered He$^*$ particles formed at any position of the interaction region is assumed to be proportional to the product of the reactant beams and the theoretical rate coefficient (obtained by multiplying the theoretical integrated cross section with the relative velocity). The elastically scattered He$^*$ particles are distributed over a Newton sphere according to the theoretically obtained DCS. The radius of this sphere is calculated from conservation of energy and momentum. The scattered products at different collision energies expand with different lab-frame velocities, and the simulation corrects for any velocity dependent detection bias in the experiment. In order to determine the final position of the particles, the scattered He$^*$ particles are propagated with their post collision lab-frame velocities from time $t$ to $t_{laser}$. All the particles that lie outside the 3 mm aperture located at the entrance of the VMI are discarded.

For evaluation of the theoretical partial relative rate coefficient, only the particles distributed over the backward hemisphere that arrive at the detection zone are counted and then normalized by the intensity of the reactant beams. Note that the laser ionization volume is estimated from the lens used to focus the laser beam. This process is repeated for every collision energy to obtain the black curve in Fig. 3. However, for the simulation of VMI images, all the particles that arrive at the detection zone are considered. The images obtained at a given collision energy are simulated by projecting 3D spheres of elastically scattered He$^*$ within the detection zone onto a 2D array with a mesh spacing of 1 m/s. The normalization and the evaluation of angular distributions for the simulated images shown in Fig. 2 is performed following the same procedure as for the experimental images.

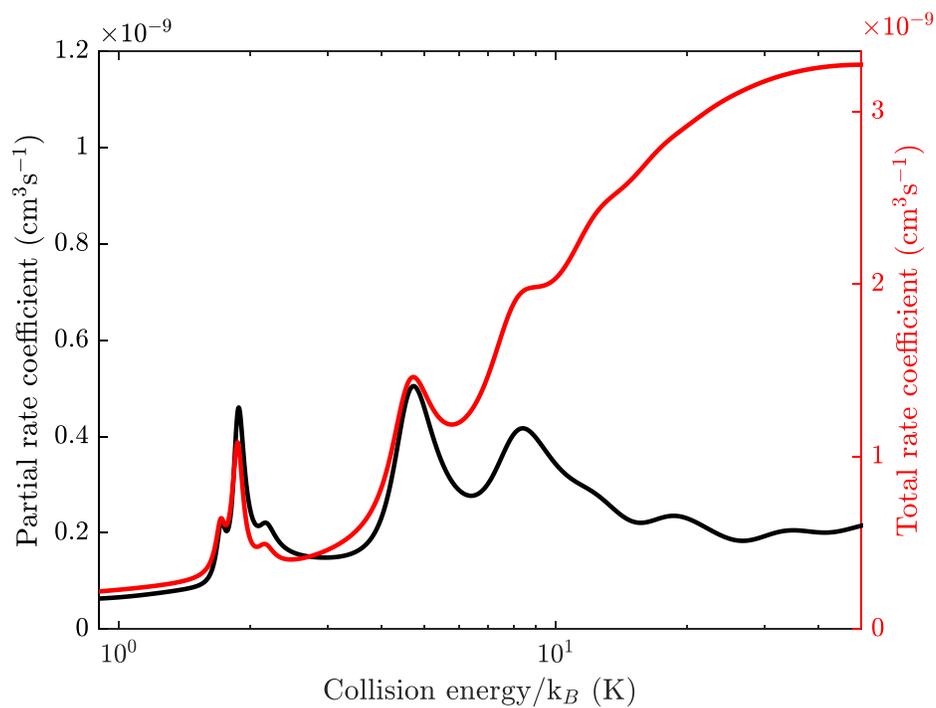

**Fig. S1 | Partial and total rate coefficients for elastic scattering of He* with normal-D$_2$.** The black curve denotes the theoretically predicted partial elastic scattering rate coefficient integrated only over the backward hemisphere. The red curve denotes the theoretically predicted total elastic scattering rate coefficient (integrated over all scattering angles). Note the different y-scales for the partial and total rate coefficients.